\documentstyle[epsfig,prd,floats,aps]{revtex}
\begin{document}
\draft

\twocolumn[\hsize\textwidth\columnwidth\hsize\csname @twocolumnfalse\endcsname

\title{An exact solution for 2+1 dimensional critical collapse}

\author{David Garfinkle\cite{dgemail}}
\address{
\centerline{Department of Physics, Oakland University,
Rochester, Michigan 48309}}

\maketitle

\null\vspace{-1.75mm}

\begin{abstract}
We find an exact solution in closed form for the critical collapse of
a scalar field with cosmological constant in 2+1 dimensions.  This
solution agrees with the numerical simulations done by Pretorius and 
Choputik\cite{mattandfrans} of this system.
\end{abstract}


\narrowtext

\vskip2pc]

\section{Introduction}

Critical gravitational collapse at the threshold of black hole formation,
as first found by Choptuik,\cite{matt} has been studied in many
systems.\cite{carsten}  With the exception of a study of vacuum,
axisymmetric collapse,\cite{evans} the systems studied are spherically
symmetric.  Because of this symmetry, the equations describing the
collapse are PDEs for functions of two variables.  The critical
solution is often discretely self-similar (DSS) or continuously
self-similar (CSS).  In the CSS case one can study the critical
solution itself by assuming the CSS symmetry and thus reducing the
collapse equations to a set of ODEs.  In general, the equations
(both the PDEs describing collapse and the ODEs describing a CSS
critical solution) are sufficiently complicated that a numerical
treatment is needed.

The Einstein equations in 2+1 dimensions are simpler than in 3+1 dimensions.
Though to form a black hole in 2+1 dimensions, one must add a cosmological
constant, this added feature still allows analytic treatment of some
collapse situations.\cite{mann}   
One might then hope that critical collapse in 2+1
dimensions would be more tractable.  Indeed, for the collapse of 
thin dust rings\cite{steif} or the collision of point 
particles\cite{cp7,birmingham}
the collapse can be treated analytically.  (This is essentially because
the spacetime has constant curvature outside of the zero thickness sources). 

Recently, Pretorius and Choptuik\cite{mattandfrans} performed
numerical simulations of the collapse of a massless, minimally coupled scalar
field with a cosmological constant in 2+1 dimensions.  They find
that the critical solution is CSS.  

In this paper, we find the critical solution of reference\cite{mattandfrans}
in closed form.  In section 2 we write the Einstein-scalar equations in
an appropriately chosen coordinate system.  In section 3 we make a CSS
ansatz and find the solution.  This solution is compared in section 4
to the numerical results of reference\cite{mattandfrans} and perturbations
of the solution are considered in section 5.

\section{Field equations}

The Einstein-scalar equations with cosmological constant are
\begin{equation}
{G_{ab}}+\Lambda {g_{ab}} = \kappa {T_{ab}}
\end{equation}
Here, $G_{ab}$ is the Einstein tensor, $T_{ab}$ is the stress energy of
the scalar field $\phi$
\begin{equation}
{T_{ab}} = {\nabla _a} \phi {\nabla _b}\phi - {1 \over 2} {\nabla ^c}\phi
{\nabla _c} \phi {g_{ab}}
\end{equation}
and $\kappa$ is a constant.  Following the conventions of
reference\cite{mattandfrans} we chose units such that $\kappa = 4 \pi$.

We now consider an appropriate choice of coordinate system for the metric.
Since we want to study the CSS critical solution, we want a coordinate
system in which the solution appears manifestly CSS.  This is not the 
case for the coordinates of reference\cite{mattandfrans} .  Instead
we use the method of Christodoulou\cite{christodoulou} to choose a
coordinate system where the coordinates are geometric quantities. 
We choose as a radial coordinate $\bar r$ such that $2 \pi {\bar r}$
is the length of the circles of symmetry.  This is the analog of the
usual area coordinate used in spherical symmetry.  We choose as our
``time'' coordinate, the null coordinate $u$ defined as follows:
$u$ is constant on outgoing light rays, and on the world line of
the central observer, $u$ is equal to the proper time of that observer.
Finally, we choose a coordinate $\theta$ so that $\partial /\partial \theta$
is the Killing vector.  The metric then takes the form
\begin{equation}   
d{s^2} = - {e^{2 \nu}} d {u^2} - 2 {e^{\nu + \lambda}} d u d {\bar r}
+ {{\bar r}^2} d {\theta ^2}
\end{equation}
where $\nu$ and $\lambda$ are functions of $u$ and $\bar r$.  In 
reference\cite{chop6} the Einstein-scalar equations were found for
spherical symmetry in any number of dimensions.  For our purposes,
we specialize the results of reference\cite{chop6} to 2+1 dimensions,
generalize them to add a cosmological constant and change the 
convention to $\kappa =4\pi$.   We begin by introducing null vectors
$l^a$ and $n^a$ defined by
\begin{equation}
{l^a} = {e^{- \lambda }} {{\left ( {\partial \over {\partial {\bar r}}}
\right ) }^a}
\end{equation}
\begin{equation}
{n^a} = {e^{- \nu }} {{\left ( {\partial \over {\partial u}}
\right ) }^a} - {1 \over 2}
{e^{- \lambda }} {{\left ( {\partial \over {\partial {\bar r}}}
\right ) }^a}
\end{equation}
Then the Einstein-scalar equations with cosmological constant are
satisfied provided that the scalar field satisfies the wave equation and that
the following components of Einstein's equation are satisfied
\begin{equation}
\label{Gll}
{G_{ab}}{l^a}{l^b} = 4 \pi {{\left ( {l^a} {\nabla _a} \phi \right ) }^2}
\end{equation}
\begin{equation}
\label{Gln}
{G_{ab}}{l^a}{n^b} = \Lambda
\end{equation}
Equations (\ref{Gll}) and (\ref{Gln}) become
\begin{equation}
\label{gfield}
{1 \over {\bar r}} {e^{- 2 \lambda }} {\partial \over 
{\partial {\bar r}}} (\lambda
+ \nu ) = 4 \pi {e^{ - 2 \lambda }} {{\left ( {{\partial \phi} \over
{\partial {\bar r}}} \right ) }^2}
\end{equation}
\begin{equation}
\label{gbarfield}
{{-1} \over {2 {\bar r}}} {e^{- 2 \lambda }} {\partial \over 
{\partial {\bar r}}} (\nu - \lambda ) = \Lambda
\end{equation}
The solution of equations (\ref{gfield}) and 
(\ref{gbarfield}) is most easily expressed by
defining the quantities $g \equiv {e^{\nu + \lambda}}$ and
${\bar g} \equiv {e^{\nu - \lambda }}$.  Then we have
\begin{equation}
\label{gcomp}
g = \exp \left [ 4 \pi {\int _0 ^{\bar r}} {\bar r} 
{{\left ( {{\partial \phi} \over
{\partial {\bar r}}} \right ) }^2} d {\bar r} \right ]
\end{equation}
\begin{equation}
\label{gbarcomp}
{\bar g} = 1 - 2 \Lambda {\int _0 ^{\bar r}} {\bar r} g d {\bar r}
\end{equation}
The wave equation for $\phi $ becomes
\begin{equation}
\label{wave}
2 {{{\partial ^2} \phi } \over {\partial u \partial {\bar r}}}
+  {1 \over {\bar r}}
{{\partial \phi } \over {\partial u}} - {1 \over {\bar r}} 
{\partial \over {\partial {\bar r}}} 
\left ( {\bar r} {\bar g} {{\partial \phi } \over 
{\partial {\bar r}}} \right ) = 0 
\end{equation}

\section{Critical solution}

We now make the ansatz that the scalar field is CSS and use this to
solve the field equations.  Choose the origin of $u$ to be at the
singularity, and define two new coordinates $T$ and $R$ by
\begin{equation}
u \equiv - {e^{ - T}} 
\end{equation}
\begin{equation}
{\bar r} \equiv {e^{ -T}} R
\end{equation}
Then demand that $\phi$ take the form
\begin{equation}
\label{ansatz}
\phi = c T + \psi (R)
\end{equation}
where $c$ is a constant.  This ansatz requires that we neglect
the cosmological constant, which in turn means that ${\bar g} =1$ and
thus reduces equation (\ref{wave}) to the flat space wave equation. 
Putting the ansatz in equation (\ref{ansatz}) into 
equation (\ref{wave}) we obtain
\begin{equation}
R (1 - 2 R) {\psi ''} + (1-3R) {\psi '} - c = 0
\end{equation}
where a prime denotes derivative with respect to $R$.  The
solution of this equation that is regular at the origin is
\begin{equation}
\psi = - 2 c \ln \left [ {1 \over 2} \left ( 1 + {\sqrt {1 - 2
R}}\right ) \right ] 
\end{equation}
which leads to a scalar field given by
\begin{equation}
\label{phisol}
\phi = c \left ( T - 2 \ln \left [ {1 \over 2} \left ( 1 + 
{\sqrt {1 - 2 R}}\right ) \right ] \right )
\end{equation}
Then using equation (\ref{gcomp}) we find that the metric function $g$
is
\begin{eqnarray}
\label{gsol}
\nonumber
g = \exp \left [ 4 \pi {\int _0 ^R} R 
{{\left ( {{\partial \phi} \over
{\partial R}} \right ) }^2} d R \right ] \\
= {{\left [ {{{(1+{\sqrt {1-2R}})}^2}
\over {4 {\sqrt{1-2R}}}}\right ] }^{8 \pi {c^2}}}
\end{eqnarray}

The metric of the CSS critical solution must be singular at $u=0$.  
However, our critical solution (equations (\ref{phisol}) and (\ref{gsol}))
appears to have an additional singularity at $R=1/2$ which is the past
light cone of the $u=0$ singularity.  We now consider whether the 
apparent singularity at $R=1/2$ is a real singularity or a coordinate
singularity.  Note that from equation (\ref{gsol}) it follows that
$R=1/2$ (for any value of $T$) is a marginally outer trapped surface
and that the Christodoulou coordinates go bad at just such surfaces.
Now define a new coordinate $v$ by 
$v=-(u+2{\bar r})$.  Then the metric is
\begin{eqnarray}
\nonumber
d {s^2} = {{\left [ {{-u} \over {16}} {{\left ( 1 + {\sqrt {v \over {-u}}}
\right ) }^4}\right ] }^{4 \pi {c^2}}} {v^{- 4 \pi {c^2}}} d v d u \\
+{1 \over 4} {{(u+v)}^2} d {\theta ^2}
\end{eqnarray}
Now define the number $q$ and the coordinate $w$ by $1/(2q) = 1 - 4 \pi {c^2}$
and ${w^{2q}}=v$.  Then the metric is 
\begin{eqnarray}
\label{smoothmetric}
\nonumber
d {s^2} = {{\left [ {{-u} \over {16}} {{\left ( 1 + {1 \over {\sqrt {-u}}}
{w^q} \right ) }^4}\right ] }^{1 - {{(2q)}^{-1}}}} 2 q d u d w \\
+ {1 \over 4} {{\left ( u + {w^{2q}}\right ) }^2} d {\theta ^2}
\end{eqnarray}
This metric is smooth at $w=0$ provided that $q=n$ where $n$ is a positive
integer.  That is, the metric is smooth for values of $c$ given by
\begin{equation}
\label{c(n)}
c = \pm {\sqrt {{1 \over {4 \pi}} \left ( 1 - {1 \over {2 n}}\right ) }}
\end{equation}
For $n=1$ the spacetime is the 2+1 dimensional Robertson-Walker metric.

We now consider the question of whether it is physically necessary to
impose the condition that the metric be smooth at $w=0$.  If $q=n$, then
one can show using equation (\ref{smoothmetric}) that the spheres of 
symmetry are trapped surfaces for $w<0$.  However, the critical solution
cannot have trapped surfaces, since it forms the boundary between those
evolutions that result in trapped surfaces and those that do not.  
Therefore, one should expect the numerical critical solution to approach
our CSS solution only inside the past light cone of the singularity
(that is for $w>0$).  It is therefore not physically necessary that
the CSS solution be smoothly extendible past $w=0$ since such an
extension cannot correspond to the behavior of the numerical critical
solution. 

\section{Comparison with numerical results}

In a near-critical collapse, the evolution at first approaches the critical
solution, and then diverges from it as the single unstable mode grows.
Therefore, in comparing a numerical simulation of near-critical collapse
to a proposed critical solution, one should make the comparison at 
an intermediate time: late enough for the critical solution to be approached,
but early enough so that the unstable mode does not have appreciable
amplitude.  

To compare the analytic and numerical results, one must express both
in the same coordinate system.  In the coordinates of
reference\cite{mattandfrans} the metric takes the form
\begin{equation}
\label{cpmetric}
d{s^2} = {{e^{2A}}\over {{\cos ^2}(r/l)}} ( d{r^2} - d {t^2}) +
{l^2} {\tan ^2} (r/l) {e^{2B}} d {\theta ^2}
\end{equation}
where $l={\sqrt {-1/\Lambda}}$ and $A$ and $B$ are functions of 
$r$ and $t$.  The coordinates of reference\cite{mattandfrans} are
related to ours by ${\bar r} = l \tan (r/l) {e^B}$ and $u$ is that
function of $t-r$ that at $r=0$ is equal to the proper time of the
central observer.
Therefore, it is fairly straightforward to take scalars and tensors in
the coordinates of reference\cite{mattandfrans} and express them
in our coordinate system.

\begin{figure}[bth]
\begin{center}
\makebox[3.0in]{\psfig{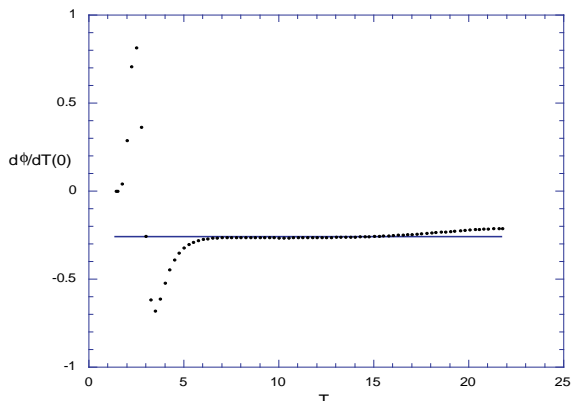}}
\caption{numerical values of ${{\left . \partial \phi /\partial T \right |
}_{R=0}}$ (dots) with constant (line) approximation to 
intermediate time behavior} 
\label{fig1}
\end{center}
\end{figure}

\begin{figure}[bth]
\begin{center}
\makebox[3.0in]{\psfig{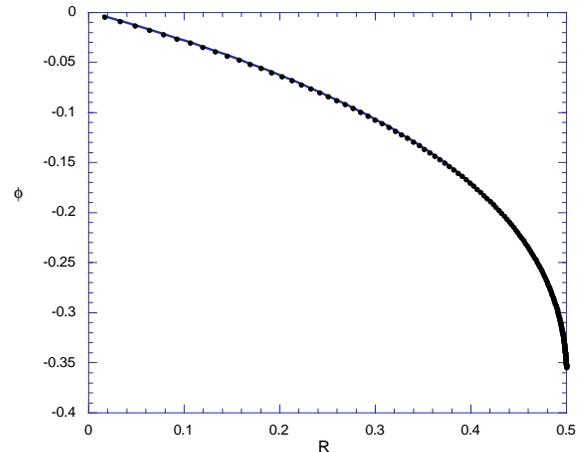}}
\caption{comparison of analytic (curve) and numerical (dots) values
of the scalar field at an intermediate time}
\label{fig2}
\end{center}
\end{figure}

Our solution has an unknown parameter $c$, 
which should be chosen for best fit with the numerical results.
To make this choice, we note that the analytical solution
(equation (\ref{phisol})) satisfies $c= {{\left . \partial \phi 
/\partial T \right | }_{R=0}}$.  Thus the values of 
${\left . \partial \phi /\partial T \right |}_{R=0}$ for the
numerical solution should allow us to determine the value of $c$.
Figure 1 shows a plot of 
${\left . \partial \phi /\partial T \right | }_{R=0} $ {\it vs} $T$
for the numerical solution.  Note that there is a range of intermediate
times for which this quantity is approximately constant.  The line
in the solution corresponds to the value of $c$ given by equation
(\ref{c(n)}) with $n=4$.  While it is not necessary that $c$ be
given by equation (\ref{c(n)}), we find that this particular value
of $c$ gives excellent agreement with the numerical data.  From now
on, we will assume that $c$ has this value.  

Figure 2 shows a comparison between the numerical and analytic
results for the scalar field $\phi$.  Here the dots are the numerical
results and the curve is the analytic one.
The comparison is made at $T=9$.  The freedom to add a constant to
the scalar field is used to set the value of the scalar field to zero
at the origin.  Note that there is excellent agreement between 
analytic and numerical results.  Though this figure shows the
comparison at only one time, the agreement persists for a large 
range of intermediate times.

\begin{figure}[bth]
\begin{center}
\makebox[3.0in]{\psfig{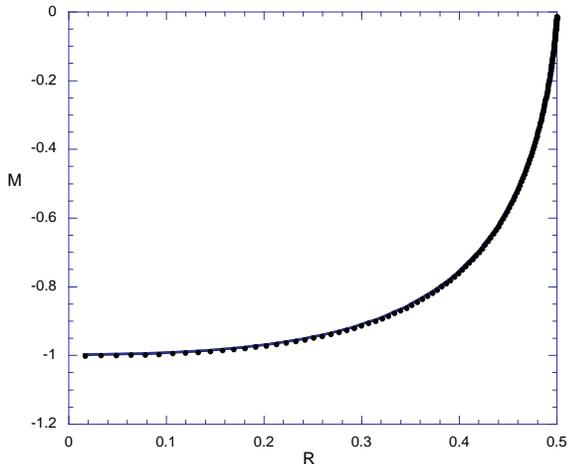}}
\caption{comparison of analytic (curve) and numerical (dots) values
of the mass aspect at an intermediate time}
\label{fig3}
\end{center}
\end{figure}

Due to the symmetry of the spactime, the metric is determined entirely
by the matter.  This is made explicit in equations (\ref{gcomp}) and 
(\ref{gbarcomp}).  Therefore, since the scalar fields of the numerical
and analytic solutions agree, the metrics must agree also.  Nonetheless,
for illustrative purposes we present a comparison of the metrics of 
the numerical and analytic solutions.  Figure 3 shows a comparison
of the quantity $M=({{\bar r}^2}/{l^2})-({\bar g}/g)$ with 
the numerical solution given
by the dots and the analytic solution given by the curve.  Note the
excellent agreement between the two solutions.  Here the comparison
is made at $T=9$, but the agreement persists for a large range of 
intermediate times.  The metric is determined by $M$ and $\bar g$
which is 1 for the analytic solution and very close to 1 for the numerical
solution in the intermediate range of times.  Thus, in the intermediate
range of times, the scalar field and metric of the numerical and analytic
solutions agree.

In figures 2 and 3, we have performed the comparison of analytic and
numerical solutions only up to the past light cone of the singularity.
However, the numerical solution certainly continues beyond this light
cone, and since $q=4$, the analytic solution also extends.  Do the two
solutions still agree in this region?  I argued at the end of the 
previous section that they cannot agree, since the analytic solution
contains a closed trapped surface beyond the light cone.  Nonetheless,
it would be useful to compare the two solutions in this larger region
to see the extent and nature of their disagreement.

To make this comparison, a different coordinate from $R$ is needed, since
the solution is not smooth in $R$ across the light cone.  Here we use
as our new coordinate the affine parameter $\lambda $ along an outgoing
radial null geodesic.  The affine parameter is made scale invariant by
choosing the null geodesic to have inner product $u$ with the central
observer.  The analytic solution can be expressed parametrically in 
terms of $\lambda $ as follows: (here we specialize to the case $q=4$).
Introduce the variable $x \equiv {{(-u)}^{-1/8}} w$.  Then $x=1$ at
the origin and $x=0$ at the past light cone of the singularity.  We
have 
\begin{equation}
\label{lam}
\lambda = {1 \over {2 {\sqrt 2}}} {\int _x ^1} {{\left ( 1 + {x^4}
\right ) }^{7/2}} d x
\end{equation}
\begin{equation}
\label{philam}
\phi = {\sqrt {7 \over {8 \pi }}} \ln \left ( {{1 + {x^4}}\over 2}\right )
\end{equation}  
\begin{equation}
\label{mlam}
M = - {x^7} {{\left ( {2 \over {1 + {x^4}}}\right ) }^{7/2}}
\end{equation}
From equation (\ref{lam}) it follows that the past light cone is at
$\lambda \approx 0.838$.  Using equation (\ref{cpmetric}), the affine 
parameter $\lambda$ can be found numerically from the numerical data.
Note, however that the numerical simulations have only a certain range
in $\lambda$ since a critical numerical solution must stop before the
time $t$ of singularity formation.  In our case, the maximum value of
$\lambda$ for the numerical data is approximately 1.   

Figures 4 and 5 show a comparison between the numerical and analytic
solutions for the range $0 \le \lambda \le 1$.  Here the values of
$\phi $ are compared in figure 4, while the mass aspect $M$ is 
compared in figure 5.  As in figures 2 and 3, the comparison is made
at $T=9$.  From the figures it is not clear whether there is a 
disagreement between the two solutions in the range $\lambda > 0.838$
that is beyond the past light cone.  To see the disagreement clearly,
one would need a larger range of $\lambda $.  This could perhaps be
provided by a numerical simulation that used double null coordinates. 

\begin{figure}[bth]
\begin{center}
\makebox[3.0in]{\psfig{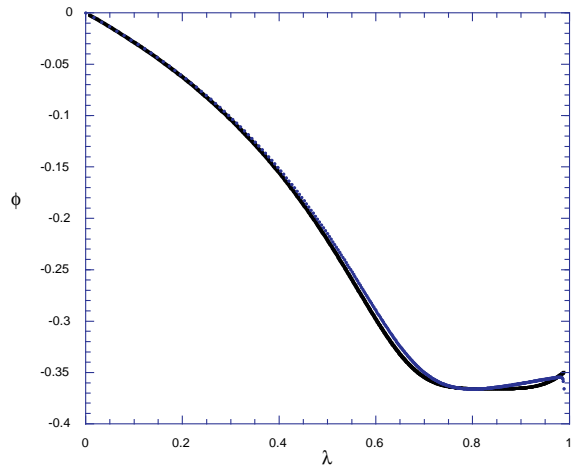}}
\caption{comparison of analytic (curve) and numerical (dots) values
of the scalar field with respect to affine parameter 
at an intermediate time}
\label{fig4}
\end{center}
\end{figure}

\begin{figure}[bth]
\begin{center}
\makebox[3.0in]{\psfig{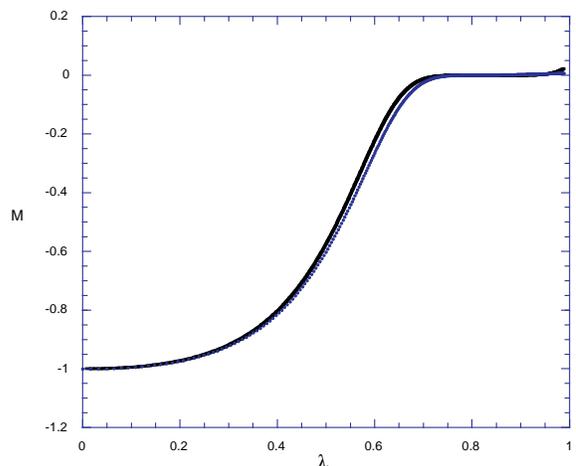}}
\caption{comparison of analytic (curve) and numerical (dots) values
of the mass aspect with respect to affine parameter
at an intermediate time}
\label{fig5}
\end{center}
\end{figure}

\section{Perturbations}

We now turn to a treatment of perturbations of the critical solution.
This treatment should in principle allow us to determine analytically
(1) the correct value of the parameter $c$ and
(2) the value of the exponent in scaling relations for near-critical
collapse.  Unfortunately, as we will see below, it is unclear what
boundary conditions to impose on the perturbations. 

The critical solution, when perturbed, has one unstable mode that grows as
$e^{kT}$ for some constant $k$.  Therefore the correct value of $c$ 
is the one for which there is exactly one unstable
mode.  The quantity $k$ is related to scaling laws for near-critical
collapse.  In the collapse of a one parameter family of initial data,
a quantity $Q$ with dimension ${({\rm length})}^s$ obeys a scaling relation
$Q \propto {{|p - {p^*}|}^{s/k}}$ where $p$ is the
parameter and $p^*$ is its critical value.  Pretorius and
Choptuik\cite{mattandfrans} examine scaling in maximum scalar curvature
for subcritical collapse, a quantity that has dimension 
${({\rm length})}^{-2}$.  They find $k \approx 0.83$.

In the approximation that ${\bar g}=1$, the scalar field satisfies the 
flat space wave equation.  Therefore, the perturbed scalar field 
$\delta \phi$ also satisfies this equation.  Then making the ansatz
$\delta \phi = {e^{kT}} S(R)$ and using equation (\ref{wave}) we obtain
\begin{equation}
R (1-2R) {S''} + (1 - [3+2k]R) {S'} - k S = 0
\end{equation}
The solution is
\begin{equation}
S = F(k,1/2,1,2R)
\end{equation}
where $F$ is a hypergeometric function.  Writing the hypergeometric
function in integral form, we have
\begin{equation}
S = {2 \over \pi} {\int _0 ^{\pi/2}} {{\left ( 1 - 2 R {\sin ^2} \chi
\right ) }^{-k}} d \chi
\end{equation}

In principle, one should now impose boundary conditions on the behavior of
the perturbation at $R=1/2$ and these boundary conditions would then
determine the allowed values of $k$.  Unfortunately, it is not clear
what boundary conditions are physically reasonable, since the critical
numerical solution should match the analytical solution only for
$R<1/2$.  Therefore, though we have the solution of the perturbation
equation, we cannot use this solution to determine the critical exponent.
     
\section{Acknowledgements}

I would like to thank Matt Choptuik and Frans Pretorius for many
helpful discussions and for providing the comparison between their
numerical work and the exact solution of this paper.  This work
was partially supported by NSF grant PHY-9722039 to Oakland University.

\end{document}